\documentclass[twocolumn,english]{revtex4}
\usepackage[T1]{fontenc}
\usepackage[utf8]{luainputenc}
\usepackage{amsmath}
\usepackage{amssymb}
\usepackage{esint}

\makeatletter
\@ifundefined{textcolor}{}
{%
 \definecolor{BLACK}{gray}{0}
 \definecolor{WHITE}{gray}{1}
 \definecolor{RED}{rgb}{1,0,0}
 \definecolor{GREEN}{rgb}{0,1,0}
 \definecolor{BLUE}{rgb}{0,0,1}
 \definecolor{CYAN}{cmyk}{1,0,0,0}
 \definecolor{MAGENTA}{cmyk}{0,1,0,0}
 \definecolor{YELLOW}{cmyk}{0,0,1,0}
}

\usepackage{babel}

\usepackage{babel}

\usepackage{babel}

\usepackage{babel}

\usepackage{babel}

\usepackage{babel}

\usepackage{babel}

\usepackage{babel}

\usepackage{babel}

\usepackage{babel}

\usepackage{babel}

\usepackage{babel}

\usepackage{babel}

\usepackage{babel}

\usepackage{babel}

\usepackage{babel}

\usepackage{babel}

\usepackage{babel}

\makeatother

\usepackage{babel}
\begin{document}

\title{Correlation functions of the Lieb-Liniger gas and the LeClair-Mussardo
formula }

\author{Garry Goldstein and Natan Andrei}

\address{Department of Physics, Rutgers University}

\address{Piscataway, New Jersey 08854}
\begin{abstract}
In this letter we derive formulas for multi point correlation functions,
in the thermodynamic limit, for the Lieb Liniger gas taken with respect
to arbitrary eigenstates. These results apply for the ground state,
thermal states and GGE states. We obtain these correlation functions
as a series of multiple integrals of progressively higher dimensions.
These integrals converge rapidly for short distance correlation functions
and low densities of particles. The series derived matches exactly
the LeClair Mussardo formula for correlation functions of relativistic
integrable models. 
\end{abstract}
\maketitle
\label{sec:Introduction}\textit{Introduction}. The field of integrable
models has reached a certain level of maturity. For many integrable
models in 1-D the exact eigenstates and eigen energies are know for
arbitrary size systems and their form in the thermodynamic limit is
well known through the string hypothesis \cite{key-31,key-27,key-29,key-40}.
The thermodynamics of many integrable models is well understood through
the thermodynamic Bethe ansatz which is also based on the string hypothesis
\cite{key-31,key-27,key-29,key-40}. One of the key remaining challenges
is the calculation of correlation functions for local observables
\cite{key-27}. This task is very important as the theoretical description
of a many body system is usually given in terms of correlation functions
of local observables. In particular the results of any experiments
that may be carried out on the system can be computed in terms of
sufficiently complicated multi point correlation functions, $G(x_{1}\cdots x_{n})=\langle O_{1}(x_{1})\cdots O_{n}(x_{n})\rangle$,
with the expectation value is taken with respect to the initial state
- typically the ground state or a thermal ensemble (which is known
to be equivalent to an eigenstate calculation \cite{key-27}). Many
correlation functions are directly measurable. For example, in the
cold atoms context, correlation functions are measurable through time
of flight interferometry \cite{key-26,key-1,key-6,key-5,key-7,key-8,key-9,key-3,key-11,key-4}.
The calculation of these correlation functions has been a major challenge
over the years and many approximation methods were developed \cite{key-27,key-29}.
Even when the system is described by an integrable Hamiltonian with
all its eigenstates given by the Bethe Ansatz \cite{key-40} or by
equivalently by the Quantum Inverse Scattering method \cite{key-27},
the correlation functions are known for only few cases and are implicitly
expressed in terms of complicated infinite determinants.

The study of correlation functions for arbitrary states, rather than
with respect to the ground state, has recently been motivated by the
study of quench dynamics in integrable systems. In some cases - when
no bound states occur and when the initial state is translationally
invariant or close to it - the system is described in the long time
limit by a generalized Gibbs ensemble (GGE) \cite{key-12,key-13,key-14,key-15,key-16,key-17,key-18,key-19,key-20,key-8-1}.
Furthermore the equilibrium state which corresponds to a GGE ensemble
is equivalent to a specially chosen eigenstate of the integrable model
\cite{key-1-1}. By judicious choice of the initial state virtually
any eigenstate may be obtained at long times. It is of great theoretical
interest to compute correlation functions for such eigenstates (or
equivalently understand quenches at long times).

The model we will study the Lieb-Liniger Hamiltonian describes bosons
moving on the continuous line and interacting via a short range potential
\cite{key-9-1}. Imposing periodic boundary condition with periodicity
$L$ the Hamiltonian is given by, $H_{LL}=\intop_{-L/2}^{L/2}dx\left\{ \partial_{x}b^{\dagger}\left(x\right)\partial_{x}b\left(x\right)+c\left(b^{\dagger}\left(x\right)b\left(x\right)\right)^{2}\right\} $,
with $b^{\dagger}(x)$ being the creation operator of the bosons at
point $x$. The $N$-bosons eigenstates of the model, $\left|k_{1}...k_{N}\right\rangle $,
labeled by rapidities $\{k_{1}...k_{N}\}\equiv\{k\}$, are explicitly
given by: 
\begin{equation}
\begin{array}{l}
\intop_{-L/2}^{L/2}dx_{N}...\intop_{-L/2}^{L/2}dx_{1}\times\\
\qquad\times\prod_{i<j}Z_{x_{i}-xj}\left(k_{i}-k_{j}\right)\prod e^{ik_{i}x_{i}}\prod b^{\dagger}\left(x_{i}\right)\left|0\right\rangle 
\end{array}\label{eq:Yudson_eigenstate}
\end{equation}
with energy $E=\sum k_{i}^{2}$ and momentum $P=\sum k_{i}$. The
scattering factor $Z_{x}\left(k\right)\equiv\frac{k+ic\left(1-2\theta\left(x\right)\right)}{k+ic}$
incorporates the $S$-matrix of the Lieb-Liniger model, $S_{LL}^{ij}=\frac{k_{i}-k_{j}-ic}{k_{i}-k_{j}+ic}$.
The $S$-matrix describes collisions between two particles in the
model and the integrability of the model hinges on the fact that products
multi-particle collisions can be consistently described in terms of
2-particles collisions \cite{key-31,key-27,key-29,key-40}. The correlation
functions in the Lieb-Liniger model take the form $\left\langle \left\{ k\right\} \right|O_{1}(x_{1})\cdots O_{n}(x_{n})\left|\left\{ k\right\} \right\rangle $
where the states $\left|\left\{ k\right\} \right\rangle $ will be
specified by the density $\rho_{p}(k)$ of the Bethe-Ansatz momenta
$k$ and the density of the holes $\rho_{h}(k)$. Both densities are
obtained from solving the Bethe-Ansatz equations that follow from
imposing periodic boundary conditions, $\exp(iLk_{j})\prod_{k\neq j}S_{LL}\left(k_{k}-k_{j}\right)=1$.
In particular all correlation functions will depend explicitly on
the occupation probability, $f\left(k\right)=\frac{\rho_{p}\left(k\right)}{\rho_{t}\left(k\right)}$,
with $\rho_{t}(k)=\rho_{p}(k)+\rho_{h}(k)$ being the total quasiparticle
density. Such states $\left|\left\{ k\right\} \right\rangle $ capture,
for appropriately chosen states both thermal averages or their GGE
generalizations which are the long time limit of the Lieb-Liniger
system quenched from some given initial state. Thus the ability to
compute expectation values is very useful in understanding the not
only the thermodynamics of this model (and other integrable models)
but also the non equilibrium evolution dynamics of integrable many
body quantum systems.

In this letter we present a series expansion formula for arbitrary
multipoint correlation functions for arbitrary eigenstates of the
Lieb Liniger gas. This expansion is given in terms of a series of
finite dimensional integrals where all the information about the eigenstate
being given in terms of the occupation probability for the state.
The series converges efficiently for a small density of particles
(or equivalently a large coupling constant) and for short distances
between the points of the correlation function. The series presented
is identical in form to the LeClair Mussardo formula \cite{key-23}
for correlation functions of relativistically invariant integrable
models. More precisely we will prove that for a generic local operator
$O$ (which can be a multipoint functions) for the Lieb liniger gas
we have that: 
\begin{align}
\left\langle \left\{ k\right\} \right|O\left|\left\{ k\right\} \right\rangle =\sum_{n}\frac{1}{n!}\int\prod_{j=1}^{n}f\left(k_{j}\right)\frac{dk_{j}}{2\pi}F_{2n}^{O}\left(k_{1},..k_{n}\right)\label{eq:LeClair_Mussardo-1}
\end{align}
Here $F_{2n}^{O}\left(k_{1},..k_{n}\right)$ is appropriately chosen
and depends only on $k_{1}...k_{n}$ (in particular there are no dressing
equations \cite{key-53,key-50-1,key-51-1,key-52-1}). We will derive
explicit formulas for $F_{2n}^{O}\left(k_{1},..k_{n}\right)$ for
short distance expansions of the field field correlation function,
for the density-density correlation function and for the emptiness
probability all of which are directly measurable in cold atoms experiments.

\textit{Algebraic Bethe ansatz}. The correlation functions are conveniently
expressed in the equivalent language of the Algebraic Bethe Ansatz
(ABA) as it applies to the Lieb Liniger model. The main object used
in the ABA is the so called transfer matrix \cite{key-27}: 
\begin{equation}
T\left(k\right)=\left(\begin{array}{ll}
A\left(k\right) & B\left(k\right)\\
C\left(k\right) & D\left(k\right)
\end{array}\right)\label{eq:Transfer_matrix}
\end{equation}
with $A\left(k\right)$, $B\left(k\right)$, $C\left(k\right)$ and
$D\left(k\right)$ are operators on the space of the bosons, e.g.
$A\left(k\right)=A\left(k,\left\{ b\left(x\right)\right\} ,\left\{ b^{\dagger}\left(y\right)\right\} \right)$
etc. Also $C\left(k\right)=B^{\dagger}\left(k\right)$. The vacuum
eigenvalues of the operators $A\left(k\right)$, $D\left(k\right)$
and $C\left(k\right)$ are given by: $C\left(k\right)\left|0\right\rangle =0,\,A\left(k\right)\left|0\right\rangle =a\left(k\right)\left|0\right\rangle ,\,D\left(k\right)=d\left(k\right)\left|0\right\rangle $,
with $a\left(k\right)=\exp\left(-i\frac{L}{2}k\right)$, $d\left(k\right)=\exp\left(i\frac{L}{2}k\right)$.
It is possible to show that the state $B\left(k_{1}\right)...B\left(k_{N}\right)\left|0\right\rangle $
is an eigenstate of the Lieb-Liniger Hamiltonian if the rapidities
$\left\{ k_{i}\right\} $ satisfy the Bethe Ansatz equations. It is
convenient to normalize our states and write: 
\[
\left|k_{1}...k_{N}\right\rangle =\left(-ic\right)^{-N/2}\prod_{j<k}\frac{1}{f\left(k_{j},k_{k}\right)}B\left(k_{1}\right)...B\left(k_{N}\right)\left|0\right\rangle 
\]
where with $f\left(k,q\right)=\frac{k-q+ic}{k-q}$. With this normalization
it is possible to show \cite{key-30} that the wavefunction of the
state $\left|k_{1},...k_{N}\right\rangle $ is given by Eq. (\ref{eq:Yudson_eigenstate})
above. Furthermore the states are normalized such that: 
\begin{equation}
\left\langle k_{1}...k_{N}\mid k_{1}....k_{N}\right\rangle =\det\left(M_{jk}\right)\label{eq:Normalization-1}
\end{equation}
with $M_{jk}=\delta_{jk}\left(L+\sum_{l=1}^{N}\frac{2c}{c^{2}+\left(k_{j}-k_{l}\right)^{2}}\right)-\frac{2c}{c^{2}+\left(k_{j}-k_{k}\right)^{2}}$,
see \cite{key-27}.

\textit{Field field correlation functions.} As an example of the general
formalism we proceed now to calculate the field field correlation
functions. Its Fourier transform yields the velocity distribution
and is directly measurable in experiment. It is known that the correlation
function for $b^{\dagger}\left(x\right)b\left(y\right)$ with respect
to a state $\left|\left\{ k\right\} \right\rangle $ with occupation
density $f\left(k\right)=\frac{\rho_{p}\left(k\right)}{\rho_{t}\left(k\right)}$
is given by \cite{key-27}: 
\begin{equation}
\frac{\left\langle \left\{ k\right\} \right|b^{\dagger}\left(x\right)b\left(y\right)\left|\left\{ k\right\} \right\rangle }{\left\langle \left\{ k\right\} \mid\left\{ k\right\} \right\rangle }=\frac{\frac{\partial}{\partial\alpha}\left(\tilde{0}\right|\det\left(1+\frac{1}{2\pi}W_{T}\right)\left|0\right)_{\alpha=0}}{\det\left(1-\frac{1}{2\pi}K_{T}\right)}\label{eq:Q_XY_Correlation-1}
\end{equation}
with the determinant extending over all the rapidities $k_{m}$ entering
the state $\left|\left\{ k\right\} \right\rangle $ and where the
operator valued matrix is given by: 
\begin{align}
W_{T}\left(k,q\right)= & \frac{f\left(q\right)}{c}\left[t\left(k,q\right)+t\left(q,k\right)\exp\left(i\left(y-x\right)\left(q-k\right)+\right.\right.\nonumber \\
 & \left.+\Phi_{A_{2}}\left(q\right)-\Phi_{D_{2}}\left(q\right)+\Phi_{D_{2}}\left(k\right)-\Phi_{A_{2}}\left(k\right)\right)-\nonumber \\
 & -\left[t\left(k,q\right)\exp\left(i\left(y-x\right)\left(q-k\right)+\Phi_{A_{1}}\left(k\right)+\right.\right.\nonumber \\
 & \left.\left.+\Phi_{D_{1}}\left(q\right)\right)+t\left(q,k\right)\exp\left(\Phi_{A_{1}}\left(q\right)+\Phi_{D_{1}}\left(k\right)\right)\right]\nonumber \\
 & \times\exp\left(\psi_{D_{2}}\left(k\right)-\psi_{A_{1}}\left(k\right)-\Phi_{A_{1}}\left(k\right)+\right.\nonumber \\
 & \left.+\psi_{A_{2}}\left(q\right)+\psi_{D_{3}}\left(q\right)-\psi_{D_{1}}\left(q\right)-\Phi_{D_{2}}\left(q\right)\right)+\nonumber \\
 & +\alpha c\exp\left(-i\left(y-x\right)k+\psi_{D_{2}}\left(k\right)-\Phi_{A_{2}}\left(k\right)+\right.\nonumber \\
 & \left.\left.+\psi_{A_{3}}\left(q\right)+\psi_{A_{2}}\left(q\right)-\psi_{D_{1}}\left(q\right)-\Phi_{D_{2}}\left(q\right)\right)\right]\label{eq:W_T-K-Q}
\end{align}
Here $f\left(k,q\right)=\frac{k-q+ic}{k-q}$, $g\left(k,q\right)=\frac{ic}{k-q}$,
$h\left(k,q\right)=\frac{k-q+ic}{ic}$ and $t\left(k,q\right)=\frac{\left(ic\right)^{2}}{\left(k-q\right)\left(k-q+ic\right)}$.
The defining relations for these fields are given by: $\Phi_{A_{k}}=P_{A_{k}}+Q_{A_{k}},\,\Phi_{D_{k}}=P_{D_{k}}+Q_{D_{k}}$
and $\psi_{A_{k}}=p_{A_{k}}+q_{A_{k}},\,\psi_{D_{k}}=p_{D_{k}}+q_{D_{k}}$.
The various expectation values are $P_{a}\left|0\right)=p_{a}\left|0\right)=\left(\tilde{0}\right|q_{a}=\left(\tilde{0}\right|Q_{a}=0,\,\left(\tilde{0}\mid0\right)=1$.
Here $\left|0\right)$ is not related to any states of the boson system
but is a fictitious auxiliary ground state. The only nonzero commutation
relations are given by: $\left[P_{A_{i}},Q_{A_{k}}\right]=\delta_{ik}\ln\left(h\left(q,k\right)\right)$,
$\left[P_{D_{i}},Q_{D_{k}}\right]=\delta_{ik}\ln\left(h\left(k,q\right)\right)$,
$\left[p_{A_{i}},q_{A_{k}}\right]=\delta_{ik}\ln\left(h\left(q,k\right)\right)$
and $\left[p_{D_{i}},q_{D_{k}}\right]=\delta_{ik}\ln\left(h\left(k,q\right)\right)$.
Furthermore the determinant $W_{T}\left(k,q\right)$ in Eq. (\ref{eq:W_T-K-Q})
is well defined since $\left[\Phi_{a}\left(k\right),\Phi_{b}\left(q\right)\right]=\left[\psi_{a}\left(k\right),\psi_{b}\left(q\right)\right]=\left[\Phi_{a}\left(k\right),\psi_{b}\left(q\right)\right]=0$.
Using the Taylor series formula for Fredholm determinants \cite{key-35}
it is possible to obtain a series solution for the determinant: 
\begin{equation}
\begin{array}{l}
\frac{\partial}{\partial\alpha}\left(\tilde{0}\right|\det\left(1+\frac{1}{2\pi}W_{T}\left(k,q\right)\right)\left|0\right)=\\
=\sum_{n=0}^{\infty}\frac{1}{n!}\int\frac{dk_{1}}{2\pi}...\int\frac{dk_{n}}{2\pi}\prod f\left(k_{j}\right)\frac{\partial}{\partial\alpha}\det\left(\tilde{0}\right|\left(\hat{W}_{T}\left(k_{j},k_{k}\right)\right)_{n\times n}\left|0\right)
\end{array}\label{eq:determinant expansion-1}
\end{equation}
Here $\left(\hat{W}_{T}\left(k_{j},k_{k}\right)\right)_{n\times n}$
is the $n\times n$ matrix whose $j,k$'th entry is $\hat{W}_{T}\left(k_{j},k_{k}\right)$,
where $\hat{W}_{T}\left(k,q\right)=\frac{W_{T}\left(k,q\right)}{f\left(q\right)}$.
Similarly we have that 
\begin{equation}
\begin{array}[t]{l}
\det\left(1-\frac{1}{2\pi}K_{T}\right)=\\
=\sum_{n=0}^{\infty}\frac{-1^{n}}{n!}\int\frac{dk_{j}}{2\pi}\prod f\left(k_{j}\right)\det\left(\frac{2c}{\left(k_{j}-k_{k}\right)+c^{2}}\right)
\end{array}\label{eq:DET_K_T-1}
\end{equation}
Hence we obtain a LeClair Mussardo like formula for the correlation
function by Taylor expanding the ratio of these determinants, \begin{widetext}
\[
\frac{\left(\tilde{0}\right|\det\left(1+\frac{1}{2\pi}W_{T}\right)\left|0\right)}{\det\left(1-\frac{1}{2\pi}K_{T}\right)}=\sum_{n=0}^{\infty}\frac{1}{n!}\int\prod f\left(k_{j}\right)\frac{dk_{j}}{2\pi}F_{2n}^{b^{\dagger}\left(x\right)b\left(y\right)}\left(k_{1},..k_{n}\right)
\]
where the $F_{2n}^{b^{\dagger}\left(x\right)b\left(y\right)}\left(k_{1},..k_{n}\right)$
are given by 
\begin{align*}
F_{2}^{b^{\dagger}\left(x\right)b\left(y\right)} & =\frac{\partial}{\partial\alpha}\det\left(\tilde{0}\right|\left(\hat{W}_{T}\left(k_{j},k_{j}\right)\right)_{1\times1}\left|0\right)\\
F_{4}^{b^{\dagger}\left(x\right)b\left(y\right)} & =\frac{\partial}{\partial\alpha}\det\left(\tilde{0}\right|\left(\hat{W}_{T}\left(k_{j},k_{k}\right)\right)_{2\times2}\left|0\right)+\frac{\partial}{\partial\alpha}\det\left(\tilde{0}\right|\left(\hat{W}_{T}\left(k_{j},k_{j}\right)\right)_{1\times1}\left|0\right)\cdot\left(\hat{K}_{T}\left(k_{k},k_{k}\right)\right)_{1\times1}
\end{align*}
etc and where $\left(\hat{K}_{T}\left(k_{j},k_{k}\right)\right)_{n\times n}$
is the $n\times n$ matrix whose $j,k$'th entry is $\hat{K}_{T}\left(k_{j},k_{k}\right)=\frac{2c}{\left(k_{j}-k_{k}\right)+c^{2}}$.
We can calculate the first few terms in the expansion for the field
field correlation function, it is given by: 
\begin{align}
\left\langle b^{\dagger}\left(x\right)b\left(y\right)\right\rangle _{2} & =\int dk\frac{f\left(k\right)}{2\pi}\exp\left[-i\left(y-x\right)k\right]dk+\int dk\frac{f\left(k\right)}{2\pi}\int dq\frac{f\left(q\right)}{2\pi}\frac{2}{c}\exp\left[-i\left(y-x\right)k\right]+\nonumber \\
 & -\frac{1}{c}\int dk\frac{f\left(k\right)}{2\pi}\int dq\frac{f\left(q\right)}{2\pi}\exp\left[-i\left(y-x\right)q\right]\left(g\left(k,q\right)\left(h\left(q,k\right)h^{-1}\left(k,q\right)+h^{-1}\left(k,q\right)h^{-1}\left(k,q\right)+2h^{-1}\left(q,k\right)\right)\right)+\nonumber \\
 & +i\int dk\frac{f\left(k\right)}{2\pi}\int dq\frac{f\left(q\right)}{2\pi}\exp\left[-i\left(y-x\right)q\right]\left(-2i\left(x-y\right)f^{-1}\left(k,q\right)+\frac{2}{ic}h^{-1}\left(q,k\right)+\frac{4i}{c}f^{-1}\left(k,q\right)\right)+....\label{eq:Field_field_second_order_correlations}
\end{align}
\end{widetext}

From this expression we see that the LeClair Mussardo formula is a
short distance expansion (in particular the term $\sim x-y$ diverges
at large distances). This is a general feature of the expansion. This
expression is valid for $x>y$, while for $x<y$ one needs to replace
the expression by its complex conjugate. In the supplement we give
some examples of how to use these formulas for various initial states.

\textit{Density density correlation functions.} We now proceed to
consider density density correlation functions, this would be another
completely explicit example. Actually it is easier to consider the
generating function of density density correlation $\exp\left(\alpha Q_{xy}\right)$,
where $Q_{xy}=\int_{x}^{y}b^{\dagger}\left(z\right)b\left(z\right)dz$.
It is known that the correlation function for $\exp\left(\alpha Q_{xy}\right)$
with respect to a state $\left|\left\{ k\right\} \right\rangle $
with occupation density $f\left(k\right)=\frac{\rho_{p}\left(k\right)}{\rho_{t}\left(k\right)}$
is given by \cite{key-27}: 
\begin{equation}
\frac{\left\langle \left\{ k\right\} \right|\exp\left(\alpha Q_{xy}\right)\left|\left\{ k\right\} \right\rangle }{\left\langle \left\{ k\right\} \mid\left\{ k\right\} \right\rangle }=\frac{\left(\tilde{0}\right|\det\left(1+\frac{1}{2\pi}V_{T}\right)\left|0\right)}{\det\left(1-\frac{1}{2\pi}K_{T}\right)}\label{eq:Q_XY_Correlation}
\end{equation}
Here the we have the ratio of two Fredholm determinants. $K_{T}\left(k,q\right)=\frac{2c}{\left(k-q\right)^{2}+c^{2}}f\left(q\right)$,
and 
\begin{align}
V_{T}\left(k,q\right)= & \frac{f\left(q\right)}{c}\left[t\left(k,q\right)+t\left(q,k\right)\exp\left(-i\left(y-x\right)\left(k-q\right)\right)\times\right.\nonumber \\
 & \times\exp\left\{ \varphi_{1}\left(q\right)-\varphi_{1}\left(k\right)\right\} +e^{\alpha}\exp\left\{ \varphi_{3}\left(k\right)+\right.\nonumber \\
 & \left.+\varphi_{4}\left(q\right)\right\} \times\left(t\left(q,k\right)+t\left(k,q\right)\times\right.\nonumber \\
 & \left.\left.\exp\left(-i\left(y-x\right)\left(k-q\right)\right)\exp\left\{ \varphi_{2}\left(k\right)-\varphi_{2}\left(q\right)\right\} \right)\right]\label{eq:V_T_k_q}
\end{align}
The various fields are defined as $\varphi_{i}\left(k\right)=q_{i}\left(k\right)+p_{i}\left(k\right),\,i=1,2,3,4$.
The commutation relations for these fields are given by $\left[p_{a}\left(k\right),p_{b}\left(q\right)\right]=\left[q_{a}\left(k\right),q_{b}\left(q\right)\right]=0$
and $\left[p_{a}\left(k\right),q_{b}\left(q\right)\right]=H_{a,b}\left(k,q\right)$.
With $H_{a,b}\left(k,q\right)=\left(\ln\left(h\left(k,q\right)\right)\right)A+\left(\ln\left(h\left(q,k\right)\right)\right)A^{T}$,
\[
A=\left(\begin{array}{cccc}
-1 & 0 & 0 & -1\\
0 & -1 & 1 & 0\\
1 & 0 & -1 & 1\\
0 & -1 & 1 & -1
\end{array}\right).
\]
The vacuum states $\left|0\right)$ and $\left(\tilde{0}\right|$
are defined as $p_{a}\left(k\right)\left|0\right)=\left(\tilde{0}\right|q_{a}\left(k\right)=0,\,\left(\tilde{0}\mid0\right)=1$.
Furthermore this determinant is well defined since $\left[\varphi_{i}\left(k\right),\varphi_{j}\left(q\right)\right]=0,\,i,j=1,2,3,4$.
Proceeding with a Taylor expansion much like in the field field case
we obtain that the density density correlation function is given by:\begin{widetext}
\begin{align}
\left\langle \rho\left(x\right)\rho\left(y\right)\right\rangle  & =-\frac{1}{2}\frac{d^{2}}{d\alpha^{2}}\frac{d^{2}}{dxdy}\left\langle \exp\left(\alpha Q_{xy}\right)\right\rangle =\nonumber \\
 & =\rho^{2}+\frac{1}{c^{2}}\int dk\int dq\frac{f\left(k\right)}{2\pi}\frac{f\left(q\right)}{2\pi}\frac{d^{2}}{d\alpha^{2}}\frac{d^{2}}{dxdy}\left(\tilde{0}\right|\left\{ 2e^{\alpha}2t\left(q,k\right)t\left(k,q\right)\exp\left(-i\left(y-x\right)\left(k-q\right)\right)\times\right.\nonumber \\
 & \times\exp\left(\varphi_{1}\left(k\right)-\varphi_{1}\left(q\right)\right)\exp\left(\varphi_{3}\left(q\right)+\varphi_{4}\left(k\right)\right)+e^{2\alpha}t^{2}\left(k,q\right)\exp\left(-i\left(y-x\right)\left(k-q\right)\right)\times\nonumber \\
 & \left.\times\exp\left(\varphi_{3}\left(k\right)+\varphi_{4}\left(q\right)\right)\exp\left(\varphi_{2}\left(k\right)-\varphi_{2}\left(q\right)\right)\exp\left(\varphi_{3}\left(q\right)+\varphi_{4}\left(k\right)\right)\right\} \left|0\right)+....=\nonumber \\
 & =\rho^{2}-\int dk\int dq\frac{f\left(k\right)}{2\pi}\frac{f\left(q\right)}{2\pi}\exp\left(-i\left(y-x\right)\left(k-q\right)\right)+...\label{eq:Correlator_Density_density}
\end{align}

\end{widetext}We see that to leading order in the small density expansion
we have that $\left\langle \rho\left(x\right)\rho\left(y\right)\right\rangle \cong\rho^{2}-\left|\left\langle b^{\dagger}\left(x\right)b\left(y\right)\right\rangle _{1}^{2}\right|$
\cite{key-42}. For example this means that at zero temperature: $\left\langle \rho\left(x\right)\rho\left(y\right)\right\rangle =\rho^{2}-\frac{\sin^{2}\left(\left(y-x\right)k_{F}\right)}{\pi^{2}\left(y-x\right)^{2}}+...$.

\textit{Probability of forming an empty interval.} Of particular interest
interest is the expectation value of the operator $\left\langle \exp\left(\alpha Q_{xy}\right)\right\rangle $
in the limit when $\alpha\rightarrow-\infty$. In that case the only
terms that contribute to the expectation value are those when there
are no particles in the interval $\left[x,y\right]$ and $Q_{xy}=0$.
We call this expectation value the probability of having an empty
interval. In this limit the expression given in Eq. (\ref{eq:Q_XY_Correlation})
greatly simplifies \cite{key-27}: 
\begin{align}
P\left(x,y\right) & \equiv\frac{\left\langle \left\{ k\right\} \right|\exp\left(\alpha Q_{xy}\right)_{\alpha\rightarrow-\infty}\left|\left\{ k\right\} \right\rangle }{\left\langle \left\{ k\right\} \mid\left\{ k\right\} \right\rangle }\label{eq:Q_XY_Emptiness}\\
 & =\frac{\left(0\right|\det\left(1+\frac{1}{2\pi}M_{T}\right)\left|0\right)}{\det\left(1-\frac{1}{2\pi}K_{T}\right)}
\end{align}

Here 
\begin{align}
M_{T}\left(k,q\right) & =-cf\left(q\right)\left[\frac{\exp\left(\frac{i}{2}k\left(y-x\right)+\frac{1}{2}\phi\left(k\right)\right)}{\left(k-q\right)\left(k-q+ic\right)}\times\right.\nonumber \\
 & \times\exp\left(-\frac{i}{2}q\left(y-x\right)-\frac{1}{2}\phi\left(q\right)\right)+\nonumber \\
 & +\frac{\exp\left(\frac{i}{2}q\left(y-x\right)+\frac{1}{2}\phi\left(q\right)\right)}{\left(q-k\right)\left(q-k+ic\right)}\times\nonumber \\
 & \left.\times\exp\left(-\frac{i}{2}k\left(y-x\right)-\frac{1}{2}\phi\left(k\right)\right)\right]\label{eq:M-matrix}
\end{align}

Here $\phi\left(k\right)=P\left(k\right)+Q\left(k\right)$ with $P\left(k\right)\left|0\right)=\left(0\right|Q\left(k\right)=0=\left[P\left(k\right),P\left(q\right)\right]=\left[Q\left(k\right),Q\left(q\right)\right]$
and $\left[P\left(k\right),Q\left(q\right)\right]=\ln\left(\frac{c^{2}}{\left(k-q\right)^{2}+c^{2}}\right)$.
The leading order expression for this correlation function (keeping
only terms with $2\times2$ matrices or less and noticing that $P\left(x=y\right)=1$)
is given by:\begin{widetext} 
\begin{align*}
P\left(x,y\right) & =1-\int\frac{f\left(k\right)}{2\pi}\left(y-x\right)+\frac{1}{2}\int\frac{f\left(k\right)}{2\pi}\frac{f\left(q\right)}{2\pi}\left(y-x\right)^{2}\\
 & -\frac{1}{2}\int\frac{f\left(k\right)}{2\pi}\frac{f\left(q\right)}{2\pi}\left\{ \frac{\left(k-q\right)^{2}+c^{2}}{\left(k-q\right)^{2}}\left(\frac{\exp\left(i\left(y-x\right)\left(q-k\right)\right)-1}{\left(q-k+ic\right)^{2}}+\frac{\exp\left(i\left(y-x\right)\left(k-q\right)\right)-1}{\left(k-q+ic\right)^{2}}\right)\right\} 
\end{align*}

For the correlation functions with respect to the ground state, $f\left(k\right)=\theta\left[-k_{F},k_{F}\right]$,
and we obtain that: 
\begin{align*}
P\left(x,y\right) & =1-\frac{k_{F}}{\pi}\left(y-x\right)+\frac{k_{F}^{2}}{\pi^{2}}\left(y-x\right)^{2}-\frac{1}{2}\int_{-1}^{1}\int_{-1}^{1}\frac{1}{4\pi^{2}}\left\{ \frac{\left(k-q\right)^{2}+\left(\frac{c}{k_{F}}\right)^{2}}{\left(k-q\right)^{2}}\times\right.\\
 & \times\left.\left(\frac{\exp\left(i\left[k_{F}\left(y-x\right)\right]\left(q-k\right)\right)-1}{\left(q-k+i\left(\frac{c}{k_{F}}\right)\right)^{2}}+\frac{\exp\left(i\left[k_{F}\left(y-x\right)\right]\left(k-q\right)\right)-1}{\left(k-q+i\left(\frac{c}{k_{F}}\right)\right)^{2}}\right)\right\} 
\end{align*}

For small $\left(y-x\right)$ this series gives $P\left(x,y\right)=1-\frac{k_{F}}{\pi}\left(y-x\right)+\frac{k_{F}^{2}}{\pi^{2}}\left(y-x\right)^{2}\left\{ \left(\frac{3}{2}-\frac{c}{k_{F}}\arctan\left(\frac{2k_{F}}{c}\right)+\left(\frac{c^{2}}{k_{F}^{2}}\right)\ln\left(1+\frac{4k^{2}}{c^{2}}\right)\right)\right\} $.

\end{widetext}

\label{sec:General-correlation-functions}\textit{General correlation
functions.} We would like to extend our results to general multipoint
correlation functions. The most general correlation function can be
written as: 
\begin{equation}
O=\prod_{j=0}^{n-1}\phi_{0,\pm1}\left(x_{j}\right)\exp\left(\theta_{j}Q_{x_{j},x_{j+1}}\right)\label{eq:General_operator}
\end{equation}

Here $\phi_{0}\left(x_{j}\right)=1$, $\phi_{1}\left(x_{j}\right)=b^{\dagger}\left(x_{j}\right)$,
$\phi_{-1}\left(x_{j}\right)=b\left(x_{j}\right)$ and $Q_{x_{j},x_{j+1}}=\lim_{\epsilon\downarrow0}\int_{x_{j}+\epsilon}^{x_{j+1}+\epsilon}b^{\dagger}\left(y\right)b\left(y\right)dy$.
Furthermore let us denote by $\alpha_{0}$ the set of $x_{j}$ where
we have a $\phi_{0}\left(x_{j}\right)$; $\alpha_{1}$ the set of
$x_{j}$ where we have $b^{\dagger}\left(x_{j}\right)$ and $\alpha_{-1}$
the set of $x_{j}$ where we have $\phi_{-1}\left(x_{j}\right)$.
Let the cardinality of $\alpha_{\pm1}$ be equal to $m$. In the supplementary
online information we show that: 
\begin{equation}
\frac{\left\langle \left\{ k\right\} \right|O\left|\left\{ k\right\} \right\rangle }{\left\langle \left\{ k\right\} \mid\left\{ k\right\} \right\rangle }=\frac{\left(\tilde{0}\right|\frac{\partial}{\partial\beta_{1}}...\frac{\partial}{\partial\beta_{m}}\det\left(1+\frac{1}{2\pi}\hat{U}_{T}\right)_{\beta_{1}=..=\beta_{m}=0}\left|0\right)}{\det\left(1-\frac{1}{2\pi}K_{T}\right)}\label{eq:Main_relation}
\end{equation}
Where $\hat{U}_{T}\left(k,q\right)\sim f\left(k\right)$. Arguing
again exactly in the same way as following Eq. (\ref{eq:determinant expansion-1})
we can obtain a series expansion for a general correlation function
thereby providing efficient methods to evaluate them a deriving the
LeClair mussardo formula.

\label{sec:Conclusions}\textit{Conclusions.} Here we have presented
a LeClair Mussardo like formula for generic correlations of the Lieb-Liniger
gas. We have proven that for a generic local operator $O$ we have
the result that is given in Eq. (\ref{eq:LeClair_Mussardo-1}). We
have presented a useful expansion for correlation functions of the
Lieb liniger gas at short distances and low densities. This result
can be used in the future to compute correlation functions for quench
problems. We hypothesis that a similar relation is valid for arbitrary
integrable models. In these models there there are many particle types.
We conjecture the following formula for the correlation functions
\begin{align}
\left\langle \left\{ k\right\} \right|O\left|\left\{ k\right\} \right\rangle = & \sum_{n}\frac{1}{n!}\int\frac{dk_{1}}{2\pi}...\int\frac{dk_{n}}{2\pi}\times\nonumber \\
 & \times\sum_{p_{i}}\left(\prod_{j=1}^{n}f^{p_{j}}\left(k_{j}\right)\right)F_{2n}^{O,p_{1}...p_{n}}\left(k_{1},..k_{n}\right)\label{eq:LeClair_Mussardo-1-1-1}
\end{align}
Here $\sum_{p_{i}}$ is a sum over particle types. It is of importance
to derive this relation for correlation functions for a general integrable
models. The authors are also currently working on extending the formula
to multi time correlation functions.

\textbf{Acknowledgments}: This research was supported by NSF grant
DMR 1410583 and Rutgers CMT fellowship.

\part*{Supplementary online information}

\section{\label{sub:LeClair-Mussardo-formula}LeClair Mussardo formula}

We begin by reviewing some properties of relativistically invariant
integrable one dimensional systems (for simplicity we focus on the
single species of particle case- generalizations to multiple species
of particles is straightforward). Eigenstates of integrable models
are parametrized by sets of rapidities $\left|\left\{ \theta_{i}\right\} \right\rangle $.
For finite sized systems with periodic boundary conditions these rapidities
satisfy bethe ansatz equations 
\begin{equation}
e^{ip\left(\theta_{j}\right)L}\prod_{k\neq j}S\left(\theta_{j}-\theta_{k}\right)=1\label{eq:bethe_ansatz_equation}
\end{equation}

Here $L$ is the system size, $p\left(\theta\right)=m\sinh\left(\theta\right)$
is the momentum of the particle (here $m$ is the mass of the particle)
and $S\left(\theta_{j}-\theta_{k}\right)$ is the scattering matrix
between particles $j$ and $k$. In integrable models the two particle
scattering matrix determines all multiparticle scattering. In the
thermodynamic limit when both the particle number and the system size
is large it is possible to introduce quasiparticle densities. To do
so let us denote by $L\rho_{p}\left(\theta\right)d\theta$ as the
number of particles in the interval $\left[\theta,\theta+d\theta\right]$,
$L\rho_{h}\left(\theta\right)d\theta$ as the number of holes in the
interval $\left[\theta,\theta+d\theta\right]$ and $L\rho_{t}\left(\theta\right)d\theta$
as the number of states in the interval $\left[\theta,\theta+d\theta\right]$
so that $\rho_{t}\left(\theta\right)=\rho_{p}\left(\theta\right)+\rho_{h}\left(\theta\right)$.
It is also convenient to introduce $f\left(\theta\right)\equiv\frac{\rho_{p}\left(\theta\right)}{\rho_{t}\left(\theta\right)}$.
The quasiparticle density satisfies the so called thermodynamic Bethe
ansatz equations: 
\begin{equation}
\rho_{t}\left(\theta\right)=\frac{1}{2\pi}p'\left(\theta\right)+\int_{-\infty}^{\infty}\frac{d\theta'}{2\pi}\varphi\left(\theta-\theta'\right)\rho_{p}\left(\theta'\right)\label{eq:thermodynamic_bethe asatz}
\end{equation}

Here $\varphi\left(\theta\right)\equiv-i\frac{d}{d\theta}\log\left(S\left(\theta\right)\right)$.
This is the continuum version of Eq. (\ref{eq:bethe_ansatz_equation}).
Using these quantities various thermodynamic quantities like the Yang-Yang
entropy may be defined. The Yang-Yang entropy associated with the
densities, $\{\rho_{p}\left(\theta\right),\rho_{h}\left(\theta\right)\}$,
measures the number of states $\left|\left\{ k\right\} \right\rangle $
consistent with the densities. It is given by: 
\begin{equation}
\begin{array}{l}
S\left(\left\{ \rho^{n}\right\} \right)=\\
=\int_{-\infty}^{\infty}d\theta\left(\rho_{h}\left(\theta\right)\ln\left(\frac{\rho_{t}\left(\theta\right)}{\rho_{h}\left(\theta\right)}\right)+\rho_{p}\left(\theta\right)\ln\left(\frac{\rho_{t}\left(\theta\right)}{\rho_{p}\left(\theta\right)}\right)\right).
\end{array}\label{eq:Entorpy}
\end{equation}
With these definitions the LeClair Mussardo formula for the expectation
of an operator with respect to an eigenstate of the theory may be
written as 
\begin{align}
\left\langle \left\{ k\right\} \right|O\left|\left\{ k\right\} \right\rangle = & \sum_{n}\frac{1}{n!}\int\frac{d\theta_{1}}{2\pi}...\int\frac{d\theta_{N}}{2\pi}\times\nonumber \\
 & \times\left(\prod_{j=1}^{n}f\left(\theta_{j}\right)\right)F_{2n,c}^{O}\left(\theta_{1},..\theta_{n}\right)\label{eq:LeClair_Mussardo}
\end{align}

Here $f\left(\theta\right)=\frac{\rho_{p}\left(\theta\right)}{\rho_{t}\left(\theta\right)}$
corresponds to the state $\left|\left\{ \theta\right\} \right\rangle $
and $F_{2n,c}^{O}\left(\theta_{1},..\theta_{n}\right)$ is the so
called connected correlation functions for the operator $O$. Its
definition is given below. To define the connected correlator recall
the Kinematic pole axiom for correlation functions of integrable field
theories \cite{key-29} (on the infinite interval) which in the simplest
case says: 
\begin{equation}
\begin{array}[t]{l}
-i\lim_{\tilde{\theta}\rightarrow\theta}\left(\tilde{\theta}-\theta\right)\left\langle \tilde{\theta},\theta_{1}^{'},...\theta_{N}^{'}\mid O\mid\theta,\theta_{1},..\theta_{N}\right\rangle \\
\qquad=\left(\prod S\left(\theta_{i}^{'}-\theta\right)-\prod S\left(\theta-\theta_{i}\right)\right)\times\\
\qquad\times\left\langle \theta_{1}^{'},...\theta_{N}^{'}\mid O\mid\theta_{1},..\theta_{N}\right\rangle 
\end{array}\label{eq:Kinematic_Pole}
\end{equation}

The pole when $\tilde{\theta}\rightarrow\theta$ is universal but
the residue may be more complex then shown in Eq. (\ref{eq:Kinematic_Pole})
above. As a result of this axiom, the expectation value of $\left\langle \theta_{1}+\epsilon_{1},...\theta_{n}+\epsilon_{n}\right|O\left|\theta_{1}...\theta_{n}\right\rangle $
has poles when $\epsilon_{i}\downarrow0$. The connected part of a
correlation function is defined by $F_{2n,c}^{O}\left(\theta_{1},..\theta_{n}\right)=$
\begin{equation}
Finite\,Part\left(\lim_{\epsilon_{i}\downarrow0}\left\langle \theta_{1}+\epsilon_{1},...\theta_{n}+\epsilon_{n}\right|O\left|\theta_{1}...\theta_{n}\right\rangle \right)\label{eq:Connected_correlator}
\end{equation}

In particular it does not have any factors of the form $\frac{\epsilon_{i}}{\epsilon_{j}}$.

\section{\label{sub:Lieb-Liniger-gas-as}Lieb-Liniger gas as a non-relativistic
limit of the Sinh-Gordon model}

The LeClair Mussardo formula has been defined for relativistic field
theories. We would like to explain why it should work for the Lieb-Liniger
gas. We would like to note that it has never been rigorously proved
for general correlation functions. We closely follow the derivations
given in \cite{key-20}. In the work the authors show that the Lieb-Liniger
gas (given by the Hamiltonian in the main text) can be obtained as
the nonrelativistic limit of the Sinh Gordon model. The Sinh Gordon
model consists of a single massive boson governed by the following
Lagrangian: 
\begin{equation}
\mathcal{L}=\frac{1}{2v^{2}}\left(\frac{\partial\phi}{\partial t}\right)^{2}-\left(\frac{\partial\phi}{\partial x}\right)^{2}-\frac{v^{2}}{4g^{2}}\left(\cosh\left(g\phi\right)-1\right)\label{eq:Sinh_Gordon_Lagrangian}
\end{equation}

Here $v$ is the velocity of light. The particles in this model have
mass 
\begin{equation}
M^{2}=\frac{1}{4}\frac{\sin\left(\pi\,\alpha\right)}{\pi\alpha}\label{eq:Mass}
\end{equation}

Here $\alpha=\frac{vg^{2}}{8\pi+cg^{2}}$. The non-relativistic limit
is obtained by the following procedure 
\begin{equation}
v\rightarrow\infty,\,g\rightarrow0,\,vg=const\label{eq:Non_relativsitic_Limit}
\end{equation}

In this case it is possible to obtain, by ignoring some terms oscillating
at a frequency $\frac{1}{2}v^{2}$, the Lieb-Liniger Hamiltonian given
in the main text with the coupling constant 
\begin{equation}
c=\frac{v^{2}g^{2}}{16}\label{eq:Coupling_Constant}
\end{equation}

We note that the results in this section are not a proof of the LeClair
Mussardo formula which is merely a hypothesis for the Sinh Gordon
model, but are merely motivational for our discussion below.

\section{Multipoint Correlation Function}

We would like to extend our results to general correlation functions.
The most general correlation function can be written as: 
\begin{equation}
O=\prod_{j=0}^{n-1}\phi_{0,\pm1}\left(x_{j}\right)\exp\left(\theta_{j}Q_{x_{j},x_{j+1}}\right)\label{eq:General_operator-1}
\end{equation}

Here $\phi_{0}\left(x_{j}\right)=1$, $\phi_{1}\left(x_{j}\right)=b^{\dagger}\left(x_{j}\right)$,
$\phi_{-1}\left(x_{j}\right)=b\left(x_{j}\right)$ and $Q_{x_{j},x_{j+1}}=\lim_{\epsilon\downarrow0}\int_{x_{j}+\epsilon}^{x_{j+1}+\epsilon}b^{\dagger}\left(y\right)b\left(y\right)dy$.
Because of translation invariance we may as well assume $x_{0}=0$.
Furthermore let us denote by $\alpha_{0}$ the set of $x_{j}$ where
we have a $\phi_{0}\left(x_{j}\right)$; $\alpha_{1}$ the set of
$x_{j}$ where we have $b^{\dagger}\left(x_{j}\right)$ and $\alpha_{-1}$
the set of $x_{j}$ where we have $\phi_{-1}\left(x_{j}\right)$.
Let the cardinality of $\alpha_{\pm1}$ be equal to $m$. Now we want
to calculate 
\begin{equation}
\left\langle 0\right|C\left(k_{1}^{C}\right)...C\left(k_{N}^{C}\right)OB\left(k_{1}^{B}\right)...B\left(k_{N}^{B}\right)\left|0\right\rangle \label{eq:Expectation_value-1}
\end{equation}

Take the limit $k_{j}^{C}=k_{j}^{B}$ and express everything in terms
of a determinant given in Eq. (\ref{eq:Main_relation}) in the main
text. Now we use the notation: 
\begin{align}
T_{\left(0,L\right)}\left(k\right)\left(0,L\right) & =\prod_{j=0}^{n}T_{\left(x_{j},x_{j+1}\right)}\left(k\right)\label{eq:Transfer_matrix_Product-1}\\
 & =\prod_{j=0}^{n}\left(\begin{array}{cc}
A_{j}\left(k\right) & B_{j}\left(k\right)\\
C_{j}\left(k\right) & D_{j}\left(k\right)
\end{array}\right)
\end{align}

Here $x_{n+1}=L$ and $T_{\left(a,b\right)}\left(k\right)$ is the
transfer matrix for the interval $\left(a,b\right)$. We now use the
relationship \cite{key-27}: 
\begin{align}
B\left(k_{1}^{B}\right)...B\left(k_{N}^{B}\right)\left|0\right\rangle  & =\sum_{\left\{ k^{B}\right\} =\cup_{j=0}^{n}\left\{ k_{j}^{B}\right\} }\prod_{j=0}^{n}B_{j}\left(k_{j}^{B}\right)\left|0_{j}\right\rangle \times\nonumber \\
 & \times\prod_{0\leq j<k\leq n}a_{k}\left(k_{j}^{B}\right)d_{j}\left(k_{k}^{B}\right)f\left(k_{j}^{B},k_{k}^{B}\right)\nonumber \\
\left\langle 0\right|C\left(k_{1}^{C}\right)...C\left(k_{N}^{C}\right) & =\sum_{\left\{ k^{B}\right\} =\cup_{j=0}^{n}\left\{ k_{j}^{B}\right\} }\left\langle 0_{j}\right|\prod_{j=0}^{n}C_{j}\left(k_{j}^{C}\right)\times\nonumber \\
 & \times\prod_{0\leq j<k\leq n}d_{k}\left(k_{j}^{C}\right)a_{j}\left(k_{k}^{C}\right)f\left(k_{k}^{C},k_{j}^{C}\right)\label{eq:States_products-1}
\end{align}

Here $\left|0_{j}\right\rangle $ is the state with no bosons on the
interval $\left(x_{j},x_{j+1}\right)$. Next we use the identity \cite{key-27}:
\begin{align}
b\left(x_{j}\right)\prod_{l=1}^{n_{j}}B\left(k_{l}^{B}\right)\left|0_{j}\right\rangle  & =-i\sqrt{c}\sum_{\left\{ k_{0,j}^{B}\right\} }a_{j}\left(k_{0,j}^{B}\right)\nonumber \\
 & \times\prod_{k_{l}^{B}\neq k_{0.j}^{B}}f\left(k_{0,j}^{B},k_{l}^{B}\right)\prod_{k_{l}^{B}\neq k_{0.j}^{B}}B_{j}\left(k_{l}^{B}\right)\left|0_{j}\right\rangle \nonumber \\
\left\langle 0_{j}\right|\prod_{l=0}^{n_{j}}C_{j}\left(k_{l}^{C}\right)b^{\dagger}\left(x_{j}\right) & =i\sqrt{c}\sum_{\left\{ k_{0,j}^{C}\right\} }d_{j}\left(k_{0,j}^{C}\right)\nonumber \\
 & \times\prod_{k_{l}^{C}\neq k_{0.j}^{C}}f\left(k_{l}^{C},k_{0,j}^{C}\right)\prod_{k_{l}^{C}\neq k_{0.j}^{C}}\left\langle 0\right|C_{j}\left(k_{l}^{C}\right)\label{eq:operators_product_states-1}
\end{align}

Next we use the relationship \cite{key-27}: 
\begin{equation}
\begin{array}{l}
\left\langle 0_{j}\right|\prod_{l=0}^{n_{j}}C_{j}\left(k_{l}^{C}\right)\prod_{l=1}^{n_{j}}B\left(k_{l}^{B}\right)\left|0_{j}\right\rangle =\\
=\prod_{j>k}g\left(k_{j}^{C},k_{k}^{C}\right)g\left(k_{k}^{B},k_{j}^{B}\right)\left(0\right|\det_{n_{j}}\left(S^{j}\right)\left|0\right)
\end{array}\label{eq:Scalar_product-1}
\end{equation}

Here 
\begin{align}
S_{lm}^{j} & =t\left(k_{l}^{C},k_{m}^{B}\right)a_{j}\left(k_{l}^{C}\right)d_{j}\left(k_{m}^{B}\right)\exp\left(\Phi_{A_{j}}\left(k_{l}^{C}\right)+\Phi_{D_{j}}\left(k_{m}^{B}\right)\right)+\nonumber \\
 & +t\left(k_{m}^{B},k_{l}^{C}\right)a_{j}\left(k_{m}^{B}\right)d_{j}\left(k_{l}^{C}\right)\exp\left(\Phi_{A_{j}}\left(k_{m}^{B}\right)+\Phi_{D_{j}}\left(k_{l}^{C}\right)\right)\label{eq:S-J-ML-INNER_PRODUCT-1}
\end{align}

Here 
\begin{equation}
\Phi_{A_{j}}\left(k\right)=Q_{A_{j}}\left(k\right)+P_{D_{j}}\left(k\right)\:\Phi_{D_{j}}=Q_{D_{j}}\left(k\right)+P_{A_{j}}\left(k\right)\label{eq:Field_definitions-3}
\end{equation}

The only nonzero commutation relations are given by: 
\begin{equation}
\begin{array}{l}
\left[P_{D_{j}}\left(k\right),Q_{D_{l}}\left(q\right)\right]=\delta_{jl}\ln\left(h\left(k,q\right)\right)\\
\left[P_{A_{j}}\left(k\right),Q_{A_{l}}\left(q\right)\right]=\delta_{jl}\ln\left(h\left(q,k\right)\right)
\end{array}\label{eq:Commutation_relations-1-2}
\end{equation}

The relevant expectation values are given by: 
\begin{equation}
P_{a}\left|0\right)=\left(0\right|Q_{a}=0\label{eq:expectation_values-2}
\end{equation}

We note that the state $\left|0\right)$ is not related to the state
with no bosons in any way but is merely an auxiliary vacuum used for
the purpose of calculating correlation functions. Once again 
\begin{equation}
\left[\Phi_{a}\left(k\right),\Phi_{b}\left(q\right)\right]=0\label{eq:Zero_commuatators-1}
\end{equation}

With these results we can see that:\begin{widetext} 
\begin{align*}
\left\langle 0\right|C\left(k_{1}^{C}\right)...C\left(k_{N}^{C}\right)OB\left(k_{1}^{B}\right)...B\left(k_{N}^{B}\right)\left|0\right\rangle  & =\left(-1\right)^{\left[P_{B}\right]+\left[P_{C}\right]}\prod_{j>k}g\left(k_{j}^{C},k_{k}^{C}\right)g\left(k_{k}^{B},k_{j}^{B}\right)\sum_{\left\{ k^{B}\right\} =\cup_{j=0}^{n}\left\{ k_{j}^{B}\right\} \cup_{\left\{ \alpha_{-1}\right\} }\left\{ k_{j0}^{B}\right\} }\times\\
 & \times\sum_{\left\{ k^{C}\right\} =\cup_{j=0}^{n}\left\{ k_{j}^{C}\right\} \cup_{\left\{ \alpha_{1}\right\} }\left\{ k_{j0}^{C}\right\} }\times\prod_{0\leq j<k\leq n}a_{k}\left(k_{j}^{B}\right)d_{j}\left(k_{k}^{B}\right)h\left(k_{j}^{B},k_{k}^{B}\right)\times\\
 & \times\prod_{0\leq j<k\leq n}d_{k}\left(k_{j}^{C}\right)a_{j}\left(k_{k}^{C}\right)h\left(k_{k}^{C},k_{j}^{C}\right)\times\prod_{j}\left(0\right|\det_{n_{j}}\left(\tilde{S}^{j}\right)\left|0\right)\times\\
 & \times\prod_{\left\{ \alpha_{-1}\right\} }\sqrt{c}a_{j}\left(k_{0,j}^{B}\right)\prod_{k_{l}^{B}\neq k_{0.j}^{B}}h\left(k_{0,j}^{B},k_{l}^{B}\right)\times\prod_{\left\{ \alpha_{1}\right\} }\sqrt{c}d_{j}\left(k_{0,j}^{C}\right)\prod_{k_{l}^{C}\neq k_{0.j}^{C}}h\left(k_{l}^{C},k_{0,j}^{C}\right)
\end{align*}

\end{widetext} where $\left[P_{B}\right]$ is the parity of the permutation
$\left\{ 1,2,3,...N\right\} \rightarrow\cup_{j=0}^{n}\left(k_{j0}^{B}\cup\left\{ k_{j}^{B}\right\} \right)$
and $\left[P_{C}\right]$ is the parity of the permutation $\left\{ 1,2,3,...N\right\} \rightarrow\cup_{j=0}^{n}\left(k_{j0}^{C}\cup\left\{ k_{j}^{C}\right\} \right)$
, and $\tilde{S}_{lm}^{j}=e^{\theta_{j}}S_{lm}^{j}$. Now introducing
$\Gamma_{j}=\sum_{l\in\left\{ \alpha_{1}\right\} ,l\leq j}1-\sum_{l\in\left\{ \alpha_{-1}\right\} ,l\leq j}1$.
We have that 
\begin{equation}
\begin{array}[t]{l}
\left\langle 0\right|C\left(k_{1}^{C}\right)...C\left(k_{N}^{C}\right)OB\left(k_{1}^{B}\right)...B\left(k_{N}^{B}\right)\left|0\right\rangle =\\
\prod_{j>k}g\left(k_{j}^{C},k_{k}^{C}\right)g\left(k_{k}^{B},k_{j}^{B}\right)\frac{\partial}{\partial\beta_{1}}...\frac{\partial}{\partial\beta_{m}}\left(0\right|\det S\left|0\right)_{\beta_{1}=...=\beta_{m}=0}
\end{array}\label{eq:Main_determinant-2}
\end{equation}

To define $S$ we need the following functional fields 
\begin{equation}
\begin{array}[t]{l}
\Phi_{A_{j,k}}\left(k\right)=Q_{A_{j,k}}\left(k\right)+P_{D_{j,k}}\left(k\right)\:\Phi_{D_{j,k}}=Q_{D_{j,k}}\left(k\right)+P_{A_{j,k}}\left(k\right)\\
\Psi_{A_{j,k}}\left(k\right)=\tilde{Q}_{A_{j,k}}\left(k\right)+\tilde{P}_{D_{j,k}}\left(k\right)\:\Psi_{D_{j,k}}=\tilde{Q}_{D_{j,k}}\left(k\right)+\tilde{P}_{A_{j,k}}\left(k\right)
\end{array}\label{eq:Functional fields_j_k-1}
\end{equation}

Here $0\leq j<k\leq n$. The nonzero commutation relations are given
by: 
\begin{equation}
\begin{array}[t]{l}
\left[P_{D_{j,k}}\left(k\right),Q_{D_{l,m}}\left(q\right)\right]=\delta_{(j,k),(l,m)}\ln\left(h\left(k,q\right)\right)\\
\left[P_{A_{j,k}}\left(k\right),Q_{A_{lm}}\left(q\right)\right]=\delta_{(j,k),(l,m)}\ln\left(h\left(q,k\right)\right)\\
\left[\tilde{P}_{D_{j,k}}\left(k\right),\tilde{Q}_{D_{l,m}}\left(q\right)\right]=\delta_{(j,k),(l,m)}\ln\left(h\left(k,q\right)\right)\\
\left[\tilde{P}_{A_{j,k}}\left(k\right),\tilde{Q}_{A_{lm}}\left(q\right)\right]=\delta_{(j,k),(l,m)}\ln\left(h\left(q,k\right)\right)
\end{array}\label{eq:Commutation_relations-1-1-1}
\end{equation}

The relevant expectation values are given by: 
\begin{equation}
P_{a}\left|0\right)=\left(0\right|Q_{a}=\tilde{P}_{a}\left|0\right)=\left(0\right|\tilde{Q}_{a}=0\label{eq:expectation_values-1-2}
\end{equation}

Furthermore we will need the functional fields: 
\begin{equation}
\begin{array}[t]{l}
\psi_{A_{j}}\left(k\right)=q_{A_{j}}\left(k\right)+p_{D_{j}}\left(k\right)\:\psi_{D_{j}}=q_{D_{j}}\left(k\right)+p_{A_{j}}\left(k\right)\\
\varphi_{A_{k}}\left(k\right)=r_{A_{k}}\left(k\right)+s_{D_{k}}\left(k\right)\:\varphi_{D_{k}}=r_{D_{k}}\left(k\right)+s_{A_{k}}\left(k\right)
\end{array}\label{eq:Field_definitions-1-1}
\end{equation}

Here $j\in\alpha_{-1}$, $k\in\alpha_{1}$. The non-zero commutation
relations are given by: 
\begin{equation}
\begin{array}[t]{l}
\left[p_{D_{j}}\left(k\right),q_{D_{l}}\left(q\right)\right]=\delta_{jl}\ln\left(h\left(k,q\right)\right)\\
\left[p_{A_{J}}\left(k\right),q_{A_{l}}\left(q\right)\right]=\delta_{jl}\ln\left(h\left(q,k\right)\right)\\
\left[s_{D_{j}}\left(k\right),r_{D_{l}}\left(q\right)\right]=\delta_{jl}\ln\left(h\left(k,q\right)\right)\\
\left[s_{A_{J}}\left(k\right),r_{A_{l}}\left(q\right)\right]=\delta_{jl}\ln\left(h\left(q,k\right)\right)
\end{array}\label{eq:commutation_relations-1-1}
\end{equation}

The relevant expectation values are given by: 
\begin{equation}
p_{a}\left|0\right)=\left(0\right|q_{a}=s_{a}\left|0\right)=\left(0\right|r_{a}=0\label{eq:expectation_values-1-1-1}
\end{equation}

In terms of these variables 
\begin{equation}
S=\sum_{j\in\alpha_{0}}S^{j,0}+\sum_{j\in\alpha_{1}}S^{j,1}+\sum_{j\in\alpha_{-1}}S^{j,-1}+\sum_{l=1}^{m}\beta_{l}\Lambda^{l}\label{eq:main_determiant_definition-2}
\end{equation}

Here:\begin{widetext} 
\begin{align}
S_{k,l}^{j,0} & =-1^{\Gamma_{j}}\tilde{S}_{k,l}^{j}\prod_{o>j}a_{o}\left(k_{l}^{B}\right)\prod_{o<j}d_{o}\left(k_{l}^{B}\right)\prod_{o>j}d_{o}\left(k_{k}^{C}\right)\prod_{o<j}a_{o}\left(k_{k}^{C}\right)\times\nonumber \\
 & \times\exp\left(\sum_{o>j}\Phi_{A_{j,o}}\left(k_{l}^{B}\right)+\sum_{o<j}\Phi_{D_{o,j}}\left(k_{l}^{B}\right)+\sum_{o>j}\Psi_{D_{j,o}}\left(k_{l}^{C}\right)+\sum_{o<j}\Psi_{A_{o,j}}\left(k_{l}^{C}\right)\right)\label{eq:S_J_K-1}
\end{align}

Furthermore: 
\begin{align}
S_{k,l}^{j,1} & =S_{k,l}^{j,0}\times\exp\left(\psi_{A_{j}}\left(k_{k}^{C}\right)\right)\nonumber \\
S_{k,l}^{j,-1} & =S_{k,l}^{j,0}\times\exp\left(\varphi_{D_{j}}\left(k_{l}^{B}\right)\right)\label{eq:Field_definitions-2-1}
\end{align}

We now note that: 
\begin{align}
\Lambda_{k,l}^{d} & =c\prod_{o>l}a_{o}\left(k_{l}^{B}\right)\prod_{o<i}d_{o}\left(k_{l}^{B}\right)\prod_{o>i}d_{o}\left(k_{k}^{C}\right)\prod_{o<j}a_{o}\left(k_{k}^{C}\right)\times d_{j}\left(k_{k}^{C}\right)a_{i}\left(k_{l}^{B}\right)\times\exp\left(\psi_{D_{j}}\left(k_{k}^{C}\right)+\varphi_{A_{i}}\left(k_{l}^{B}\right)\right)\nonumber \\
 & \times\exp\left(\sum_{o>i}\Phi_{A_{i,o}}\left(k_{l}^{B}\right)+\sum_{o<i}\Phi_{D_{o,i}}\left(k_{l}^{B}\right)+\sum_{o>j}\Psi_{D_{j,o}}\left(k_{l}^{C}\right)+\sum_{o<j}\Psi_{A_{o,j}}\left(k_{l}^{C}\right)\right)\label{eq:Terms_main_determinant_definitions-1}
\end{align}

Where $j$ is the $d$'th entry of $\alpha_{1}$ and $i$ is the $d$'th
entry of $\alpha_{-1}$. Next we need to manipulate these formulas
a little bit, we start with: 
\begin{align}
\left(0\right|\det S\left|0\right) & =\prod_{j=1}^{N}a\left(k_{j}^{C}\right)d\left(k_{j}^{B}\right)\left(0\right|\prod_{j=1}^{N}\exp\left(\Phi_{A_{n}}\left(k_{j}^{C}\right)+\Phi_{D_{n}}\left(k_{j}^{B}\right)\right)\prod_{j<n}\exp\left(\Phi_{D_{j,n}}\left(k_{m}^{B}\right)+\Psi_{A_{j,n}}\left(k_{l}^{B}\right)\right)\det\tilde{S}\left|0\right)\nonumber \\
 & =\prod_{j=1}^{N}a\left(k_{j}^{C}\right)d\left(k_{j}^{B}\right)\prod_{j,k=1}^{N}h\left(k_{j}^{C},k_{k}^{B}\right)\left(\tilde{0}\right|\det\tilde{S}\left|0\right)\label{eq:Reorganisation-1}
\end{align}
Where 
\begin{equation}
\left(\tilde{0}\right|=\left(0\right|\prod_{j=1}^{N}\exp\left(P_{D_{n}}\left(k_{j}^{C}\right)+P_{A_{n}}\left(k_{j}^{B}\right)\right)\prod_{j<n}\exp\left(\Phi_{D_{j,n}}\left(k_{m}^{B}\right)+\Psi_{A_{j,n}}\left(k_{l}^{B}\right)\right)\label{eq:New_Vacuum-1}
\end{equation}
and $\tilde{S}_{l,m}=S_{l,m}d\left(k_{l}^{C}\right)a\left(k_{m}^{B}\right)\exp\left(-\Phi_{A_{n}}\left(k_{l}^{C}\right)-\Phi_{D_{n}}\left(k_{m}^{B}\right)\right)$.
Furthermore $\left(\tilde{0}\mid0\right)=1$. Now introducing $\tilde{\Phi}_{A_{n}}=\Phi_{A_{n}}-\left(\tilde{0}\right|\Phi_{A_{n}}\left|0\right)$
, $\tilde{\Phi}_{D_{n}}=\Phi_{D_{n}}-\left(\tilde{0}\right|\Phi_{D_{n}}\left|0\right)$
, $\tilde{\Phi}_{A_{j,n}}=\Phi_{A_{j,n}}-\left(\tilde{0}\right|\Phi_{A_{j,n}}\left|0\right)$,
$\tilde{\Psi}_{D_{j,n}}=\Psi_{D_{j,n}}-\left(\tilde{0}\right|\Psi_{D_{J,n}}\left|0\right)$
and specializing to the case when $k_{j}^{B}=k_{j}^{C}$ we obtain
that \cite{key-27}: 
\begin{align}
\tilde{S}_{lm} & =\left(\sum_{j\in\alpha_{0},j\neq n}S_{lm}^{j,0}+\sum_{j\in\alpha_{1}}S_{lm}^{j,1}+\sum_{j\in\alpha_{-1}}S_{lm}^{j,-1}+\sum_{l=1}^{m}\beta_{l}\Lambda^{l}\right)\exp\left(-\tilde{\Phi}_{A_{n}}\left(k_{l}^{B}\right)-\tilde{\Phi}_{D_{n}}\left(k_{m}^{B}\right)\right)a\left(k_{l}^{B}\right)d\left(k_{m}^{B}\right)\times\label{eq:main_determiant_definition-1-1}\\
 & \times\prod_{j<n}\exp\left(-\Phi_{D_{j,n}}\left(k_{m}^{B}\right)-\Psi_{A_{j,n}}\left(k_{l}^{B}\right)\right)+\delta_{lm}\left(L+\sum_{o=1}^{N}K\left(k_{m}^{B},k_{o}^{B}\right)\right)+\hat{S}_{lm}^{n}
\end{align}

Where the expressions for $S_{lm}^{j,0}+S_{lm}^{j,1}+S_{lm}^{j,-1}+\Lambda^{l}$
we have $\Phi_{A_{j,n}}\rightarrow\tilde{\Phi}_{A_{j,n}}$ and $\Psi_{D_{j,n}}\rightarrow\tilde{\Psi}_{D_{j,n}}$.
Here 
\begin{align}
\hat{S}_{lm}^{n} & =\left(d\left(k_{l}^{B}\right)a\left(k_{m}^{B}\right)t\left(k_{l}^{C},k_{m}^{B}\right)a_{n}\left(k_{l}^{C}\right)d_{n}\left(k_{m}^{B}\right)+a\left(k_{l}^{B}\right)d\left(k_{m}^{B}\right)t\left(k_{m}^{B},k_{l}^{C}\right)a_{n}\left(k_{m}^{B}\right)d_{n}\left(k_{l}^{C}\right)\times\right.\label{eq:Diagonal_S_N-1}\\
 & \left.\times\exp\left(\tilde{\Phi}_{A_{n}}\left(k_{m}^{B}\right)+\tilde{\Phi}_{D_{n}}\left(k_{l}^{C}\right)-\tilde{\Phi}_{D_{n}}\left(k_{m}^{B}\right)-\tilde{\Phi}_{A_{n}}\left(k_{l}^{C}\right)\right)\right)\prod_{j<n}a_{j}\left(k_{l}^{B}\right)d_{j}\left(k_{m}^{B}\right)\nonumber 
\end{align}
\end{widetext} Now we factorize out the part $\propto\delta_{lm}\left(L+\sum_{o=1}^{N}K\left(k_{m}^{B},k_{o}^{B}\right)\right)$.
First we note that in the thermodynamic limit $L+\sum_{o=1}^{N}K\left(k_{m}^{B},k_{o}^{B}\right)=2\pi L\rho_{t}\left(k_{m}^{B}\right)$.
Now introduce the matrix $\varTheta_{lm}=\delta_{lm}2\pi L\rho_{t}\left(k_{m}^{B}\right)$.
Using this matrix we can see that 
\begin{equation}
\left(\tilde{0}\right|\det\tilde{S}\left|0\right)=\det\varTheta_{lm}\left(\tilde{0}\right|\det\left(1+\frac{1}{2\pi}U_{T}\left(k,q\right)\right)\left|0\right)\label{eq:Main_determinant-1-1}
\end{equation}

Where 
\begin{equation}
U_{T}\left(k,q\right)=f\left(q\right)\tilde{S}\left(k,q\right)-\delta_{lm}\left(L+\sum_{o=1}^{N}K\left(k_{m}^{B},k_{o}^{B}\right)\right)\label{eq:Main_defining_relation-1}
\end{equation}

Now using the relationship \cite{key-27}: 
\begin{equation}
\begin{array}[t]{l}
\left\langle 0\right|C\left(k_{1}^{B}\right)...C\left(k_{N}^{B}\right)B\left(k_{1}^{B}\right)...B\left(k_{N}^{B}\right)\left|0\right\rangle \\
=\prod_{j=1}^{N}2\pi L\rho_{t}\left(k_{j}^{B}\right)\times\prod_{j,k}f\left(k_{j}^{B},k_{k}^{B}\right)\det\left(1-\frac{1}{2\pi}K_{T}\right)
\end{array}\label{eq:iNNER_Product-1}
\end{equation}

We obtain that 
\begin{equation}
\frac{\left\langle \left\{ k\right\} \right|O\left|\left\{ k\right\} \right\rangle }{\left\langle \left\{ k\right\} \mid\left\{ k\right\} \right\rangle }=\frac{\left(\tilde{0}\right|\frac{\partial}{\partial\beta_{1}}...\frac{\partial}{\partial\beta_{m}}\det\left(1+\frac{1}{2\pi}\hat{U}_{T}\right)_{\beta_{1}=..=\beta_{m}=0}\left|0\right)}{\det\left(1-\frac{1}{2\pi}K_{T}\right)}\label{eq:Main_relation-1}
\end{equation}
Arguing again exactly as in the main text we can obtain a LeClair
Mussardo like formula.

\section{\label{sec:Field-Field-correlation}Field Field correlation functions
(examples)}

\subsection{\label{sub:Field-Field-correlations}Field Field correlations}

We would like to give an example of how the various formulas to the
field field correlation functions work in practice. These formulas
are useful for calculation the short distance correlation functions
for various initial states. For example keeping the leading order
term in Eq. (\ref{eq:Field_field_second_order_correlations}) we obtain
that: 
\begin{equation}
\left\langle b^{\dagger}\left(x\right)b\left(y\right)\right\rangle _{1}=\int dk\frac{f\left(k\right)}{2\pi}\exp\left[-i\left(y-x\right)k\right]dk+...\label{eq:Field_field_simplified}
\end{equation}

We would like to evaluate this expression for various states $f\left(k\right)$.
We will concentrate on thermal states. In the limit of zero temperature
we have that $f\left(k\right)=\theta\left[-k_{F},k_{F}\right]$. From
this we obtain that 
\begin{equation}
\left\langle b^{\dagger}\left(x\right)b\left(y\right)\right\rangle _{1}\cong\frac{1}{\pi}\frac{\sin\left(\left(x-y\right)k_{F}\right)}{x-y}\label{eq:Field_Field}
\end{equation}

This is a good approximation for short distances and low densities.
For the case of a finite temperature state we have that $f\left(k\right)=\frac{1}{\exp\left(\beta\left(k^{2}-\tilde{\mu}\right)\right)+1}+O\left(\frac{1}{c^{3}}\right)$
where $\tilde{\mu}=\mu+\frac{T}{\pi}\int dq\ln\left(1+e^{-\beta\left(k^{2}-\tilde{\mu}\right)}\right)$
\cite{key-27}. From this we obtain that 
\begin{equation}
\left\langle b^{\dagger}\left(x\right)b\left(y\right)\right\rangle _{1}=\int\frac{dk}{2\pi}\frac{\exp\left[-i\left(y-x\right)k\right]}{\exp\left(\beta\left(k^{2}-\tilde{\mu}\right)\right)+1}+O\left(\frac{1}{c^{3}}\right)\label{eq:Field_field_temperature}
\end{equation}

We can consider two cases when $\tilde{\mu}<0$ and when $\tilde{\mu}>0$.
In the case when $\tilde{\mu}<0$ we obtain that: 
\begin{equation}
\left\langle b^{\dagger}\left(x\right)b\left(y\right)\right\rangle _{1}=\sum_{n=1}^{\infty}\left(-1\right)^{n-1}e^{n\beta\mu}\frac{\pi}{\sqrt{n\beta}}\exp\left(-\frac{1}{4}\frac{\left(x-y\right)^{2}}{n\beta}\right)\label{eq:Field_Field_negative_chemical.}
\end{equation}
In the case when $\tilde{\mu}>0$ we may approximate $k^{2}-\tilde{\mu}\cong\theta\left(k\right)\left(k-\sqrt{\tilde{\mu}}\right)2\sqrt{\tilde{\mu}}-\theta\left(-k\right)\left(k+\sqrt{\tilde{\mu}}\right)2\sqrt{\tilde{\mu}}$.
Using this expression we obtain that: 
\begin{align}
\left\langle b^{\dagger}\left(x\right)b\left(y\right)\right\rangle _{1} & =2Re\left[\frac{1}{i\left|x-y\right|+\beta\sqrt{\tilde{\mu}}}\times\right.\label{eq:Hypergeometric}\\
 & \left.\times\left._{2}\right.F\left._{1}\right.\left\{ 1,1-i\frac{\left|x-y\right|}{\beta\sqrt{\tilde{\mu}}},2+i\frac{\left|x-y\right|}{\beta\sqrt{\tilde{\mu}}},-e^{-\beta\mu}\right\} \right]
\end{align}
where $\left._{2}\right.F\left._{1}\right.$ is the hypergeometric
function.

\subsection{\label{sub:Velocity-probability-distributio}Velocity probability
distribution}

The velocity probability density is an easily measurable experimentally
relevant quantity \cite{key-5}. It is given by: 
\begin{equation}
P\left(v\right)\sim\int dx\int dy\exp\left(i\frac{v}{2}\left(y-x\right)\right)\left\langle b^{\dagger}\left(x\right)b\left(y\right)\right\rangle \label{eq:Velocity_probability-1}
\end{equation}

The factor of $\frac{1}{2}$ in the above formula comes from the fact
that the mass of the Lieb Liniger gas is normalized to $\frac{1}{2}$.
For the simpler case where $f\left(k\right)=f\left(-k\right)$ Eq.
(\ref{eq:Field_field_second_order_correlations}) simplifies for both
$x>y$ and $x<y$ giving:\begin{widetext} 
\begin{align}
\left\langle b^{\dagger}\left(x\right)b\left(y\right)\right\rangle _{2} & =\int dk\frac{f\left(k\right)}{2\pi}\exp\left[-i\left(y-x\right)k\right]dk+\int dk\frac{f\left(k\right)}{2\pi}\int dq\frac{f\left(q\right)}{2\pi}\frac{2}{c}\exp\left[-i\left(y-x\right)k\right]+\nonumber \\
 & -\frac{1}{c}\int dk\frac{f\left(k\right)}{2\pi}\int dq\frac{f\left(q\right)}{2\pi}\exp\left[-i\left|y-x\right|q\right]\left(g\left(k,q\right)\left(h\left(q,k\right)h^{-1}\left(k,q\right)+h^{-1}\left(k,q\right)h^{-1}\left(k,q\right)+2h^{-1}\left(q,k\right)\right)\right)+\nonumber \\
 & +i\int dk\frac{f\left(k\right)}{2\pi}\int dq\frac{f\left(q\right)}{2\pi}\exp\left[-i\left|y-x\right|q\right]\left(-2i\left|y-x\right|f^{-1}\left(k,q\right)+\frac{2}{ic}h^{-1}\left(q,k\right)+\frac{4i}{c}f^{-1}\left(k,q\right)\right)+....\label{eq:Field_field_second_order_correlations-1}
\end{align}
Now using the relations: 
\begin{align*}
\int dx\exp\left(-ik\left|x\right|+i\frac{v}{2}x\right) & =-P.V.\frac{i}{k-\frac{v}{2}}-P.V.\frac{i}{k+\frac{v}{2}}+\pi\delta\left(\frac{v}{2}-k\right)+\pi\delta\left(\frac{v}{2}+k\right)\\
\int dx\left|x\right|\exp\left(-ik\left|x\right|+i\frac{v}{2}x\right) & =P.V.\frac{1}{\left(k+\frac{v}{2}\right)^{2}}-P.V.\frac{1}{\left(k-\frac{v}{2}\right)^{2}}-i\frac{d}{dk}\delta\left(\frac{v}{2}-k\right)+i\frac{d}{dk}\delta\left(\frac{v}{2}+k\right)
\end{align*}
We get 
\begin{align}
P\left(v\right)_{2} & \sim\frac{1}{2}f\left(\frac{v}{2}\right)+f\left(\frac{v}{2}\right)\cdot\int dk\frac{1}{c}\cdot\frac{f\left(k\right)}{2\pi}+\nonumber \\
 & -\frac{1}{2c}\int dk\frac{f\left(k\right)}{2\pi}\cdot f\left(\frac{v}{2}\right)\left(g\left(k,\frac{v}{2}\right)\left(h\left(\frac{v}{2},k\right)h^{-1}\left(k,\frac{v}{2}\right)+h^{-1}\left(k,\frac{v}{2}\right)h^{-1}\left(k,\frac{v}{2}\right)+2h^{-1}\left(\frac{v}{2},k\right)\right)\right)+\nonumber \\
 & +i\int dk\frac{f\left(k\right)}{2\pi}\left(2\frac{d}{dv}\left(f\left(\frac{v}{2}\right)f^{-1}\left(k,\frac{v}{2}\right)\right)+f\left(\frac{v}{2}\right)\frac{1}{ic}h^{-1}\left(\frac{v}{2},k\right)+\frac{2i}{c}f^{-1}\left(k,\frac{v}{2}\right)\right)+c.c+P.V......\label{eq:Field_field_second_order_correlations-1-1}
\end{align}

Where $c.c.$ means the complex conjugate of the entire expression
and $P.V.$ is the principle value part. This formula becomes very
accurate for large $v$ where the principle value part becomes subleading.
For this case (if we denote by $N=\int dk\frac{f\left(k\right)}{\pi}$)
we have that 
\begin{align*}
P\left(v\right)_{2} & \sim\frac{1}{2}f\left(\frac{v}{2}\right)\left\{ 1+N\left\{ \frac{2}{c}-\frac{1}{c}g\left(0,\frac{v}{2}\right)\left(h\left(\frac{v}{2},0\right)h^{-1}\left(0,\frac{v}{2}\right)+h^{-1}\left(0,\frac{v}{2}\right)h^{-1}\left(0,\frac{v}{2}\right)+2h^{-1}\left(\frac{v}{2},0\right)\right)+\right.\right.\\
 & \left.\left.\frac{1}{c}\left(2h^{-1}\left(\frac{v}{2},0\right)-4f^{-1}\left(0,\frac{v}{2}\right)\right)\right\} \right\} +2iN\frac{d}{dv}\left(f\left(\frac{v}{2}\right)f^{-1}\left(0,\frac{v}{2}\right)\right)+c.c.
\end{align*}
\end{widetext} 
\end{document}